\documentclass[usenatbib]{mn2e}
\usepackage{graphicx}

\title[Visual extinction within 55 pc]{Local Bubble. Extinction within 55 pc?}
\author[J. Knude]{Jens Knude\thanks{E-mail: indus@nbi.ku.dk } \\
The Niels Bohr Institute, Copenhagen University, Juliane Maries Vej 30, DK-2100 K{\o}benhavn {\O}, Denmark}
\begin{document}

\date{Accepted 2013 Month 99. Received 2013 Month 99; in original form June 2013}

\pagerange{\pageref{firstpage}--\pageref{lastpage}} \pubyear{2013}

\maketitle

\label{firstpage}

\begin{abstract}

In the mapping of the local ISM it is of some interest to know where the first indications of
the boundary of the Local Bubble can be measured.
The Hipparcos distances combined to $B-V$ photometry and some sort of spectral classification
permit mapping of the spatial extinction distribution. Photometry is available for almost the complete 
Hipparcos sample and Michigan Classification is available for brighter stars south of $\delta$  =
+5 (1900). For the northern and fainter stars spectral types, e.g. the HD types, are given
but a luminosity class is often missing. The $B-V$ photometry and the parallax do, however, permit a 
dwarf/giant separation due to the value of the slope of the reddening vector compared to the gradient 
of the main sequence in a color magnitude diagram, in the form: $B-V$ $vs.$ $M_V+A_V$ = 
$V+5(1+log(\pi))$, together with the
rather shallow extinction present in the Hipparcos sample. We present the distribution of median 
$A_V(l, b)$ for stars with Hipparcos 2 distances less than 55 pc. The northern part of the first
and second quadrant has most extinction, up to $\sim$0.2 mag and the southern part of the third and 
fourth quadrant the slightest extinction, $\sim$0.05 mag. The boundary of the extinction minimum
appears rather coherent on an angular resolution of a few degrees
\end{abstract}


\section{Introduction}

A debate on the existence of non negligible amounts of material in the local interstellar
medium has taken place over several decades. The problem may be looked upon under very
different angles but could perhaps be reduced to: is any matter present and if so how much
and how is it
distributed? The local medium has observational as well as theoretical attractions. The 
distribution of the cooler, dense material has consequences for the theoretical models
as well as for the understanding of the propagation of the energetic radiation. The concept
of the solar vicinity has condensed to be apprehended as a local low density cavity in 
the ISM: the Local Bubble (or as a system of interconnected bubbles). A natural question is 
consequently how big this local bubble or low density cavity is?    

In minor photometric surveys, often for other purposes than LB studies, wall-like structures
were sometimes seen: abrupt changes of the color excess over very small distances. 
\cite{Geus1989} and \citet{jk1987} are examples showing an onset of reddening at $\approx$100 pc
in the general direction of Scorpius. \citet{Reis2011} is a more recent example where $uvby\beta$
data has been used to outline LB.

In order to indicate the size of the LB some of the parameters characterising it must be known. The
confinement of the bubble is encountered when one or more of these parameters are changed 
in a significant way. Such a parameter could be $n_H$ or the average line of sight 
reddening/extinction, \citet{Abt11}, \citet{FRS11}. For the size Abt quotes the range 
50-100 pc and Frisch, Redfield and Slavin present in the their Fig. ~4 a sky map for 
stars between 50 and 
100 pc displaying a nice coherent contour $E_{B-V}$ $\approx$ 0.1 mag confining a low
density region. The origin of the color excesses used for this map is, however not given in
any detail. In the
cavity the parameters are found in certain defining ranges, typically low density and small
extinction. The location where either of these defining parameters displays a 
rise could naively indicate the boundary of the LB. \citet{VVLR2010} and \citet{LVVPEC2013} 
used the gradient
$dE_{B-V} / dr$ = 0.0002 mag pc$^{\rm -1}$ to identify the LB rim. \citet{jk2010} used a similar
technique on calibrated 2MASS data to locate more massive clouds.

Stellar extinctions are naturally discrete measures but do have an upper distance limit. Data from
hydrogen emission in its various forms are continous but often lack precise distance estimates.

The continuity problem may partly be remedied with large stellar samples. The Hipparcos Catalogue
with its $\approx$120000 entries is a first approximation to provide continously 3D distributed
extinctions. A large fraction of this sample do have positive precise parallaxes, reasonably 
decent photometry and spectral/luminosity classification and thus offers itself for a 2D/3D
extinction study.  

\section{The Hipparcos Extinction sample}

For stars with spectral and luminosity classification it is not a problem to assign
the intrinsic color. Intrinsic colors are taken from \citet{SK1982}. If 
Schmidt-Kalers color system differs from that given in the Hipparcos 1 Catalogue the 
difference is ignored. Classification is as given in Hipparcos 1 but for the deklination
zone covered by the Michigan 5 Catalogue, \citet{hs1999}, which was not available for the
first publication of the Hipparcos Catalogue, \citet{plk1997}, it has been replaced by 
this more precise one.
For stars with a spectral type but no luminosity class we use the $(B-V)_{obs}$ $vs.$
$V_{obs}+5(1+log \pi)$ =  $M_V+A_V$ diagram to distinguish between the dwarfs and
giants. The parallax $\pi$ is taken from Hipparcos 2, \citet{VL2007}. Our first 
assumption is that any shift of a stars intrinsic position in the
$(B-V)_{0}$ $vs.$ $V_{0}+5(1+log \pi)$ diagram is caused solely by reddening/extinction.

We notice that the slope of the main sequence in the color magnitude diagram does not differ 
that much from the ratio $A_V / E_{B-V}$ and is nummerically smaller, implying that the rather 
small reddenings presumed to be 
present in the Hipparcos Catalog will not mix the dwarfs and the giants. A dividing curve
introduced in the color magnitude diagram can accordingly be used for the luminosity
separation. Subgiants are assumed to have colors identical to the dwarfs with the same
spectral type. The spectral type and the location relative to the dividing curve then provide
an estimate of the intrinsic color. Since most Hipparcos stars have 2MASS colors the giant
dwarf separation might also result from a diagram as Fig.~29 in \citet{jk2010}.

This way most stars with a positive parallax have estimated reddenings. The sample may, however,
be refined from a comparison to the comments given in the SIMBAD data base. Most of the Hipparcos
stars do have comments. We may accordingly sort out variables, stars in multiple systems, PMS
stars, stars with close companions etc. There is one important group of nearby stars that singles 
out: the high proper motion stars, many of which concentrates in a small region between the 
late main sequence and the giants in the $(B-V)_{obs}$ $vs.$ $V_{obs}+5(1+log \pi)$ diagram. 
They do accordingly not comply to the dividing curve scheme and we have left them
all out - which is unfortunate since they all are very close and would be ideally suited
for LB studies if their properties were better known. Neither are stars located in clusters 
included. Our extinction sample consists only of stars having
the star assignment in SIMBAD. The sample is reduced to some 85000 possible,
single non-variable stars without close neighbors this way.

\begin{figure} 
\includegraphics[width=88mm]{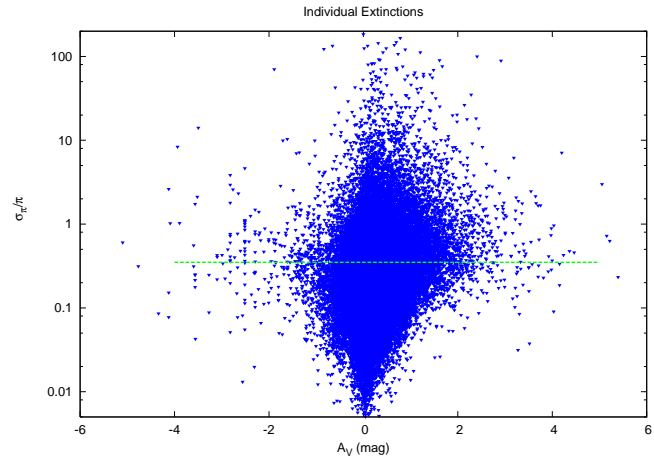}
 \caption{ $A_V$ $vs.$ $\sigma_{\pi} / \pi$ diagram for the Hipparcos extinction sample
with more than 85000 stars not commented as multiple or variable in $SIMBAD$.
High proper motion stars are not included since a large fraction of them has a strange location
in the  $(B-V)_{obs}$ $vs.$ $V_{obs}+5(1+log \pi)$ diagram.
Astrometry is from \citet{VL2007}. The change in precision with $A_V$ taking place
at $\sigma_{\pi} / \pi$ = 0.35 defines the sample used presently for the study of the
local ISM} \label{f1}
\end{figure}
 
So far we have not considered the parallax precision. In Fig.~\ref{f1} we have 
plotted $individual$ $A_V$ $vs.$ 
$\sigma_{\pi} / \pi$ where $A_V$ = 3.1$\times E_{B-V}$. Read from the bottom,
where the attractive small errors presumably for nearby stars are located,
we notice that the maximum positive extinctions increases with increasing 
$\sigma_{\pi} / \pi$ until a change takes place at $\approx$0.35. We take this 
dividing value as the upper limit for the relative precision. In fact no 
stars beyond $\sim$1 kpc has a better precision than this value. Most of the 
very negative extinctions belong to stars that has a tabular 
value $(B-V)_{obs}$ = 0.000 which obviously is not correct for the assigned spectral type. 
We have kept them in the sample though.

\begin{figure} 
\includegraphics[width=88mm]{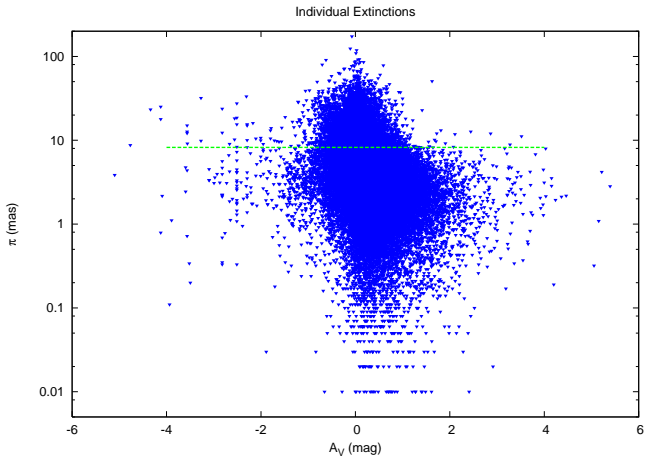}
 \caption{General variation of extinction with parallax for the Hipparcos extinction sample, 
the  $\sigma_{\pi} / \pi$ = 0.35 confinement is not applied here. Notice the extinction 
discontinuity at $\pi \approx$ 8 mas (125 pc) roughly corresponding to the canonical
"radius" of LB}
\label{f2}
\end{figure}

A diagram like $A_V$ $vs.$ $\pi$ will show the general run of extinction and should display
the LB wall, if any. There is in fact such a wall-like feature present in Fig.~\ref{f2}
where a general rise in the extinction takes place at $\pi$$\approx$8 mas or 125 pc. A
distance roughly corresponding to the maximum extent, apart from the tunnel directions,
in \citet{VVLR2010}.

\begin{figure} 
\includegraphics[width=88mm]{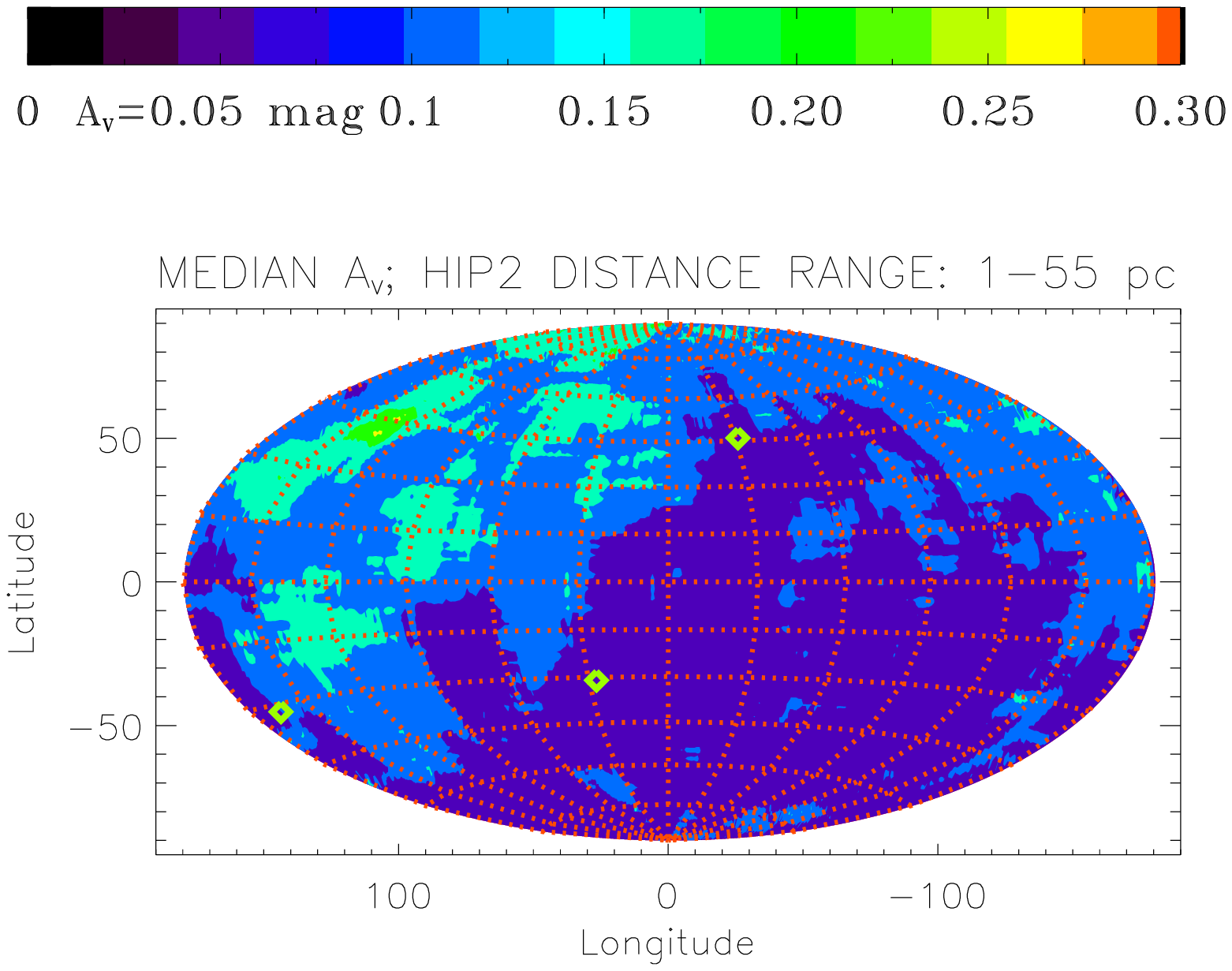}
\caption{Sky distribution of median extinction for stars within 55 pc and with 
$\sigma_{\pi} / \pi <$0.35. The location of the far UV shadow clouds, lb165-32, lb27-31 
and lb329+46 at 40, 45 and 65 pc are indicated by the diamonds, \citet{Berg1998}. 
A possible interpretation of this map could be: If the median $A_V >$ 0.1 mag the LB
boundary is less than 55 pc away and if median $A_V <$ 0.1 mag the boundary is 
farther away. Grid for every 30 degrees in longitude and for every 15 degrees 
latitude}
\label{f3}
\end{figure}

\section{Sky Map for stars within 55 pc}

On the average the Hipparcos extinction sample of $\sim$85.000 stars provide $\sim$two stars per
square degree. Of course depending on the galactic latitude. The completeness is hard to assess
in a statistical way since the Hipparcos input catalog was based on a variety of astrophysical
proposals. Since the V magnitude for completeness is between 8 and 9 we do not measure
large extinctions as is also evident from Fig.~\ref{f2}. But for our purpose, locating very nearby extinction, completeness for the
extinction values is not necessary as long as a variation defining the LB boundary can 
be detected.

We cover the sky with a grid of overlapping pixels. Pixel size determined by requesting a 
number of stars per pixel permitting the computation of a median extinction for the distance 
range under study. In this note Fig.~\ref{f3} presents the outcome for stars within 55 pc. The 
median $A_V$ values range from $\sim$0.05 mag to  $\sim$0.3 mag and we note a marked difference between
the north and south and between the 4 quadrants. This projection may be compared to Fig.~4
of \citet{FRS11} showing $E_{B-V}$ contours for stars in the distance range from 50 to 100
pc. The distribution on Fig.~\ref{f3} is almost identical but with smaller 
extinction values - which could be due to the different lenghts of the sight 
lines: $\leq$55 $vs.$ $\leq$100 pc.

In Fig.~\ref{f3} we have also indicated the location of three diffuse clouds causing 
shadows in the far UV, \citet{Berg1998}. The clouds lb165-32, lb27-31, lb329+46 are 
removed $\leq$40, 45 and 65 pc respectively, just in the volume we are studying presently.
As Fig.~\ref{f3} shows the clouds are almost superposed on the boundary of the low density
contour. Are they then part of the LB confinement? In fact they may not be because in 
\citet{Berg1998} it was found that each of these small clouds was located in front of
other clouds at 100, 135 and 125 pc respectively. These later distances agree much more to
the canonical LB confinement. Recently \citet{LVVPEC2013}, from a very large photometric 
sample of $\approx$23.000 stars with various distance indicators, have presented a $E_{B-V}$ map
for stars within 100 pc, among many others maps with different distance cuts, with a strong 
resemblence to Fig.~\ref{f3}.
So, is the LB boundary, as indicated by the rise in color excesses/extinctions, real within 
100 pc as indicated by the $E_{B-V}$ $\approx$ 0.1 mag contour, \citet{FRS11}, but not
for stars within 55 pc with an $A_V$ $\approx$ 0.1 mag contour corresponding to $E_{B-V}$ 
$\approx$ 0.03 mag as the conclusion would be if the three shadow clouds mentioned really
are $\approx$50 pc in front of the proper LB boundary? 

But if we may rely on the median extinctions and the map in Fig.~\ref{f3} and accept the median
$A_V$ = 0.1 mag as defining the onset of extinction from the LB confinement we could 
possibly state that in the northern part of the first and second quadrant the LB limit is within
55 pc and in the southern part of the third and fourth quadrant it is further away.

\section{Conclusions}
We have constructed what was termed the Hipparcos extinction sample with about 85.000 distance 
extinction pairs. The classification originally in the Hipparcos Catalogue was supplemented 
with the $Michigan \ 5 \ Catalogue$ classification of HD stars. We introduced a simple dwarf/giant
separation of the $(B-V)_{obs} \ vs. \ V_{obs}+5(1+log\pi)$ in order to have a 2D
classification of the stars with no luminosity class given in the Catalog.

Our main purpose has been to demonstrate that the Hipparcos extinction sample in addition to
estimating distances to individual clouds, e.g. \citet{jk16}, possibly also may be used to trace large scale
features, extending over large fractions of the sky. It seems justified to conclude 
that further investigation of the Hipparcos extinction sample at larger distances
may be worth while.  

\section*{Acknowledgments}

The investigation of the Milky Way ISM is supported economically by Fonden af 29. December 1967.
Simbad has been used for the extraction of the Hipparcos 1, Hipparcos 2, Michigan V catalog
data and for the comments on each of the Hipparcos stars.


\bsp

\label{lastpage}

\end{document}